# Crystallization kinetics of colloidal model suspensions: recent achievements and new perspectives

Thomas Palberg

Institut für Physik, Johannes Gutenberg Universität Mainz, 55099 Mainz, Germany

Abstract

Colloidal model systems allow studying crystallization kinetics under fairly ideal conditions with rather well characterized pair interactions and minimized external influences. In complementary approaches therefore experiment, analytic theory and simulation have been employed to study colloidal solidification in great detail. These studies were based on advanced optical methods, careful system characterization and sophisticated numerical methods. Both the effects of the type, strength and range of the pair-interaction between the colloidal particles and those of the colloid-specific polydispersity were addressed in a quantitative way. Key parameters of crystallization were derived and compared to those of metal systems. These systematic investigations significantly contributed to an enhanced understanding of the crystallization processes in general. Further, new fundamental questions have arisen and (partially) been solved over the last decade including e.g. a two step nucleation mechanism in homogeneous nucleation, choice of the crystallization pathway or the subtle interplay of boundary conditions in heterogeneous nucleation. On the other side, via the application of both gradients and external fields the competition between different nucleation and growth modes can be controlled and the resulting micro-structure be influenced. The present review attempts an account of the interesting developments occurred since the turn of the millennium and an identification of important novel trends with particular focus on experimental aspects.

**Content**

**List of symbols and abbreviations**

**Acknowledgements**

**References**

**List of Symbols and Abbreviations**

| | |
|---|---|
| $\varepsilon$ | dielectric constant |
| $\varepsilon_0$ | dielectric permittivity of vacuum |
| $\eta_P^r$ | reservoir packing fraction of polymer |
| $\Phi, \Phi_F, \Phi_M$ | packing fraction, at freezing, at melting |
| $\gamma$ | interfacial free energy |
| $\lambda$ | wavelength of light |
| $\nu$ | refractive index |
| $\sigma$ | normalized interfacial free energy |
| $\sigma_P, \sigma_{eff}$ | particle diameter, effective particle diameter |
| | |
| $a$ | radius of a colloidal particle |
| $D_S^{long}$ | long time self diffusion coefficient |
| $d_{NN}$ | nearest neighbour distance |
| $J, J^*$ | nucleation rate density, reduced nucleation rate density |
| $k_BT$ | thermal energy |
| $L$ | crystallite size |
| $N$ | Number of surface groups of a colloidal particle |
| $n$ | particle number density |
| $p_{CO}$ | coexistence pressure |
| $R$ | crystallite radius |
| $s$ | relative polydispersity |
| $X$ | fraction of crystalline material, crystallinity |
| $Z$ | bare charge of a colloidal particle |
| $Z_{eff}$ | effective charge of a colloidal particle |
| | |
| 1-EN, 2-EN | 1-ethylnaphtalene, 2-ethylnaphtalene |
| 2D, 3D | two-dimensional, three-dimensional |
| AS | attractive spheres |
| bcc | body centred cubic |
| BS | Bragg scattering |
| CS | charged spheres |
| $CO_2$ | carbon dioxide |
| fcc | face centred cubic |
| hcp | hexagonal close packed |
| HCY | hard core Yukawa |
| HS | hard spheres |
| KMJA | Kolmogorov-Mehl-Johnson-Avrami |
| NaOH | sodium hydroxide |
| Pe | Peclet number |
| PnIPAM | poly-n-isopropyl-acrylamide |
| PnBAPS | polystyrene-poly-n-butylacrylamide copolymer |
| PHSA | poly-12-hydroxy-stearic acid |
| PMMA | poly-methylmethacrylate |
| PS | polystyrene |
| SALS | small angle light scattering |
| SANS | small angle neutron scattering |
| SAXS | small angle x-ray scattering |
| USAXS | ultra small angle x-ray scattering |
| UV-VIS | ultraviolet and visible wavelength range |
| WCA | Weeks-Chandler-Anderson (potential) |

# 1. Introduction

The field of colloidal crystallization has considerably matured over the last fifty years. Its origins trace back to controversially discussed indications of a phase transition in systems of hard spheres from just introduced computer simulations [1] and to first compelling experimental observations of colloidal crystallization in industrial latex suspensions [2] and precious opal [3]. It has been experiencing substantial enhancement through fruitful application of suitable optical techniques [4] also encompassing related questions on phase behaviour and crystal properties [5, 6, 7]. It was further boosted by Pusey's colloids as atoms paradigm [8] and during the nineties studies started to address a quantitative characterization of crystallization kinetics in systems of different interaction types. The state of the art at the turn of the millennium has been summarized in Refs. [9] and [10]. Since then, further papers reviewed individual aspects like phase behaviour, nucleation and crystal growth in attractive systems [11, 12, 13], in proteins and colloids [14], in colloids and plasmas [15], in thin films [16], and in confinement [17, 18]. Furthermore, correlations between phase behaviour and solidification mechanisms [19], as well as photonic, phononic and other applications [20, 21, 22] have been reviewed. It has become clear that colloidal crystallization more than ever exerts the fascination of symmetry formation and breaking on a directly observable level [23, 24, 25, 26, 27]. Understanding the fundamentals of how crystallization proceeds under various boundary conditions established fruitful links to phase transitions in other soft matter systems [14, 15], to material design [28, 29] and to biology [30, 31, 14]. The steady increase of numerical power combined with careful system design and characterization in terms of tractable pair potentials further improved the close link between theory and experiment [32, 33, 34, 35, 36] which proved crucial for this most interesting but also challenging field.

After 15 years, the present review again focuses on the quantitative determination and interpretation of the *kinetics* of colloidal crystallization from the shear molten state and its interpretation. The review is concerned with both repulsive and attractive colloidal particles, but it will at the same time mainly restrict itself to single component systems of spherical particles, for which sufficiently accurate descriptions of the interaction and phase behaviour are available and the degree of meta-stability can be reliably estimated from theoretical considerations and/or experimentally adjusted boundary conditions.

This review will also touch some novel aspects. An important shift of interest can be seen towards the mechanism(s) of homogeneous nucleation. Combining microscopy and simulation the involved processed have been extensively studied to reveal the fascinating details of the two-step mechanism of homogeneous nucleation which comprises changes of both structure and density. Through this also the importance of kinetic pathways as opposed to thermodynamic ones has been demonstrated. Also the growth and coarsening mechanisms have attracted some interest. Like in nucleation, it was observed, that collective processes also here are more important than single particle ones.

Finally, also the now possible systematic manipulation of the crystal micro-structure will be addressed. While the crystal structure in most cases is fixed by the equilibrium phase behaviour, the micro-structure (and with it also the properties of the resulting solids) can be influenced significantly through the solidification pathway. Both heterogeneous nucleation at walls, on structured substrates and at seeds, as well as the use of additional constraints during crystallization have been applied to this end. Most prominent are attempts to crystallize the samples in confined geometry or subjected to externally applied fields and/or concentration gradients. Here the relatedness of colloids to atoms is less pronounced due to several colloid-specific properties including e.g. the low yield modules of colloidal crystals, the polydispersity of colloidal particles, the sometimes large density differences in an extended fluid-solid coexistence range in the phase diagram and the presence of a viscous solvent. Further, the presence of fields and gradients may easily lead to transport processes which destroy the initially present homogeneity and lead to interesting feed-back effects like crystallization waves or stratification. Although similar phenomena also occur in atomic systems, the details in colloidal systems often are rather system-specific and depend strongly on the applied boundary conditions. Also in discussing these novel issues, emphasis will be on experimental work, but theoretical work will be referred to, wherever suitable.

In what follows, we will first take a look at colloidal systems suitable for crystallization studies, then at developments in the available instrumentation. The main part of this review will be concerned with the homogeneous nucleation kinetics in various colloidal systems and then turn to the involved mechanisms. The next chapters will consider the possibilities of microstructure manipulation by heterogeneous nucleation and the application of external fields. The review ends with some short concluding remarks.

## 2. Colloidal systems for crystallization studies

Colloidal particles come in a large variety of mutual interactions which can be experimentally adjusted *via* synthesis and/or sample preparation [37]. Colloidal model systems for crystallization studies mainly comprise spherical particles, but also other geometries have been investigated like rods [38], ellipsoids [26] or cubes and related polygons [24, 39]. Crystallization studies are still missing for so-called colloidal molecules [40]. Quantitative crystallization studies are based mostly on repulsive hard spheres (HS) [41, 42], charged spheres (CS) [43] or spheres with attractive depletion interaction (AS) [13, 11, 44]. Other than for atomic systems, polydispersity may be a point of concern. Therefore, synthesis and sample preparation strategies aim at reducing the width of the particle size distribution [20, 45]. In general, a good model system for reference studies of crystallization kinetics in the absence of external fields qualifies by a well characterized interaction and corresponding phase behaviour as well as by some technical properties like good match of solvent and particle refractive index and mass density, facilitating multiple-scattering free optical experiments at elevated particle concentrations and studies of long duration (hours to days) without sedimentation effects, respectively.

### 2.1. Hard sphere systems and their approximants

Ideal, monodisperse HS are readily realized on a computer with perfect buoyancy match and desired packing fraction. They show a first order freezing transition [46] with a coexistence range stretching from a packing fraction of $\Phi_F = 0.492$ to $\Phi_F = 0.545$ at a coexistence pressure of $p_{co} = 11.576$ $k_BT/\sigma_P^3$, where $k_BT$ denotes the thermal energy and $\sigma_P = 2a$ is the HS diameter [47]. To put it straight, there exist no ideal HS in any experiment, but some rather close approximation can be obtained [48] after carefully performed synthesis, characterization and conditioning. Since for HS the first order freezing transition is solely driven by entropy, the determination of the sample packing fraction is crucial. Typically, sedimentation experiments are performed to fix the freezing and melting concentrations or core packing fractions [49] which then are mapped onto ideal freezing and melting packing fractions [50]. These are assumed to be known with sufficient accuracy from theory for HS, polydisperse HS of different size-distribution shape and slightly soft spheres [47, 51, 52, 53, 54, 55, 56]. Polydispersity[1] shifts the phase boundaries upward in a non trivial way and may even introduce novel phases by fractionation if it exceeds some 7-8%. Any slight softening of the potential will shift the phase boundary towards lower values. Estimates of the potential steepness can e.g. be obtained from rheological measurements and occasionally present electrostatic charges and their fluctuations from e.g. tweezing electrophoresis [57, 58, 59]. Without going into further detail, it is obvious, that ensuring well characterized experimental HS-like systems is somewhat demanding and cross-checking between different methods is necessary to facilitate unequivocal interpretation of quantitative measurements [60, 61].

---

[1] Polydispersity, s, is here characterized by the standard deviation in particle size divided by the mean particle size.

A system often employed for crystallization studies are poly-methylmethacrylate (PMMA) spheres coated with a thin layer of poly-12-hydroxystearic acid (PHSA) [62, 63]. Suspended in a mixture of cis-decalin and tetraline their contrast in refractive index may be excellently matched and the Christiansen effect [64] minimized. This allows optical investigations even at high packing fraction, avoids systematic accumulation of charges and renders their mutual interaction very close to that of HS [58]. In fact, index matching also strongly reduces the van der Waals attraction. PMMA systems are prone to sedimentation effects, which in some cases may afford µ-gravity to assure isotropic, field free conditions [65].

An alternative, charge free HS system comprises of polystyrene (PS) micro-gel particles suspended in 1- or 2-ethyl-naphtalene (1-EN, 2-EN) [66, 67]. In this good solvent of nearly identical refractive index *and* mass density, the particles remain uncharged and swell to their maximum size governed by the degree of cross-linking. The hardness of interaction can e.g. be probed in oscillatory rheological measurements [68]. For a typical degree of cross-linking of 1:30 and assuming an inverse power potential $U(r) = 1/r^n$, one typically finds values of $n \approx 40$ giving a good approximation to a HS potential. Polydispersities typically range between 0.03 and 0.07. PS micro-gel particles therefore became another often used HS model system.

Buoyancy match can also be achieved employing a mixture of cis-decalin and either bromocyclohexane or bromocycloheptane [69, 70] at the cost of charging the particles up and altering the phase behaviour towards CS [71]. In fact, due to the low dielectric constant, very large screening lengths are obtained and crystals with lattice constants of about 20µm were stabilized [25]. Addition of the organic salt tetrabutyl ammonium chloride may screen the resulting electrostatic repulsion efficiently. Other examples of slightly soft spheres are systems based on poly-N-Isopropylacrylamide (PnIPAM) micro-gel particles [72, 73, 74]. This material displays a lower solubility gap in water and shrinks with increasing temperature with a transition range somewhat above room-temperature. Here, in addition to residual charges from synthesis, also the degree of cross-linking becomes an additional control parameter [75]. In principle, if particle softness were negligible, electrostatic charges were sufficiently screened, and no attractive interactions were present, this would allow to study crystallization in HS-like system, but it is a rarely solved challenge to reach and maintain such conditions [76]. If their softness is larger, their phase behaviour becomes more complicated due to the effects of compressibility, which in addition can be deliberately combined with electrostatic repulsion by using co-polymer particles [77, 78] or altering the ratio of screening length to particle size [79]. Electrostatics may be made dominant at vanishing electrolyte concentrations [80]. Recently another slightly soft system was introduced consisting of micelles which *increase* in size upon heating. This system crystallizes upon heating [81] and like PnIPAM is suited for temperature ramp experiments.

### 2.2. *Charged spheres*

In most common theoretical studies charged spheres are modelled to interact *via* hard core Yukawa (HCY) or Debye-Hückel type pair potential, which can be adjusted in surface potential $\psi_0$ and screening length $\kappa$ by tuning the particle charge, the particle number density and the electrolyte concentration [82]. Combinations of these parameters can be expressed in terms of an effective temperature and a coupling parameter. The position of a universal melt-ing line in the effective temperature - coupling parameter plane of the phase diagram is known from computer simulations [83, 84, 85] but also other representations of the CS phase behaviour have been given, e.g. in the $\Phi$ - $1/\kappa a$ plane for both constant charge and constant potential conditions [86, 87,[88]. To obtain well characterized experimental CS systems a combination of experiments is necessary. Sizes and size distributions are available from different methods ranging from ultracentrifugation to electron microscopy [89]. Due to the long ranged nature of the CS repulsion, surface roughness or softness as well as short ranged van der Waals attractions can in most cases be neglected. The crucial parameter for the interaction strength, rather than $\Phi$, is the particle number density, $n$. Adjustment of $n$ and salt concentration, $c$, is readily facilitated employing a closed conditioning circuit [90, 91, 92] and controlled *in situ* by combining static light scattering with conductivity and/or elasticity measurements [93, 94]. For most polymer latex particles the number of charge generating surface groups, $N$, is fixed by synthesis and readily accessible by titration experiments [95]. Some systems are charged via absorption of charged surfactants and thus can be de-charged by treatment with ion exchange resin [96]. Systems with weakly acidic surface groups, however, may show incomplete dissociation and the bare charge, $Z$, may be varied by the addition of a base like NaOH [97, 98] or pyridine [99].

Of greater relevance than the bare charge is the effective charge of CS, which is regulated by several mechanisms. In physico-chemical equilibrium, only a fraction of the number of surface groups, $N$, is dissociated. Therefore the colloidal particle's bare charge, $Z$, is smaller than $N$. In crystallization experiments, the residual electrolyte concentration, $c$, is kept very low (typically below one µmol/l). Here a HCY interaction using the bare charge yields a very good description of the interaction between an isolated pair of CS. It fails, however, as soon as a third particle is considered [100]. For bulk situations, two different many body effects prevail. One is the accumulation of counter-ions close to the macro-ion surfaces due to the osmotic pressure provided by the counter-ions present in the overlapping double layer of neighbouring particles. This effect has been termed counter-ion condensation and is captured in charge renormalization theory [101, 102, 103, 82]. In addition to calculations with fixed charge or potential, also numerical studies including dissociation equilibrium (charge regulation) are available [104 and references therein].

The second effect is due to the presence of the highly charged neighbouring macro-ions themselves. These shield the repulsion effectively at the nearest-neighbour distance, $d_{NN}$ [33]. Also macro-ion shielding affords the presence of three or more particles [105], is most prominent at sufficiently long ranged HCY-interactions [106] and strongly influences the elastic properties [107]. Interestingly, quite accurate estimates of both types of effective charges can be obtained from experiment. Effective charges determined from conductivity experiments in the fluid or crystalline phase are numerically very close to theoretically calculated renormalized charges [108, 109, 110, 111]. Effective charges including many body interactions are accessible from measurements of the elasticity in the crystalline phase [93, 94] or the static structure factor, $S(q)$, of the fluid phase [112]. The effective charge, $Z_{eff}$, obtained from an interpretation of elasticity measurements in terms of an effective HCY potential has been combined with the results from computer simulations to accurately predict the phase behaviour of CS systems [113]. This was recently also confirmed for a charge variable system [98] and for a CS mixture [114] with spindle type phase diagram. This consistency between crystal properties and phase behaviour forms a solid basis for estimates of the system meta-stability from a set of independently characterized experimental boundary conditions and growth experiments [115, 10].

Except for per-fluorinated polymer spheres, CS usually show refractive indices considerably larger than that of the standard solvent water. At elevated packing fractions, multiple scatter-ing usually becomes an issue. Use of mixed solvents, e.g. glycerol/water or water/dimethyl-sulfoxide may reduce multiple scattering. In most cases, however, thoroughly deionized CS crystallize already at number densities of, say, 1/µm or even lower corresponding to packing fractions of less than one percent [116, 117, ], where multiple scattering is still negligible. Mixed solvents also may improve density match which is of relevance for per-fluorinated systems [118] and silica spheres, which also have been studied under µ-gravity [119]. Fortunately, even at low concentration CS tend to crystallize rather quickly on time scales of seconds to minutes, such that sedimentation effects usually are much less pronounced than for HS and AS. However, sealing the sample against air-borne $CO_2$ is important at such low concentrations [120, 121].

Systems used for crystallization studies mostly comprise commercially available Polymer spheres, silica particles and charged PMMA particles. In addition to close packed face centred cubic (fcc) structures the CS phase diagram also shows regions of the more open body centred cubic (bcc) structure at low particle and electrolyte concentrations [9, 98, 116, 122].

### 2.3. Attractive Spheres

Practically all colloidal spheres show a van der Waals attraction, which is typically overcome by either steric stabilization (HS) or electrostatic stabilization (CS) such that the effective potentials are rendered to a good approximation purely repulsive. If, however, non-adsorbing polymer or another small second component is added to the suspension, the so called depletion interaction results, which typically is attractive with a range proportional to the size of the added depletant and a strength proportional to its concentration [5]. Theoretically, systems of uncharged, depletion-attractive particles are often described in terms of the Asakura-Oosawa-Vrij potential [123, 124]. Here, one finds that the fluid-solid coexistence range is broadened and the coexistence pressure is considerably lowered as compared to HS. To give an idea, for a size ratio of 0.15 and a polymer reservoir packing fraction of $\eta_P^r = 0.1$ Monte Carlo simulations find $\Phi_F = 0.494$ to $\Phi_F = 0.64$ at a coexistence pressure of $p_{co} = 8.00 \, k_BT/\sigma_P^3$ [47]. Moreover, the phase diagram of monodisperse hard sphere – polymer mixtures becomes very rich. In addition to fluid and fcc phases known from HS, it reveals also a liquid phase, an attractive glass, gelation and the possibility of forming coexisting meta-stable fluid and/or crystal phases [125, 126, 127]. In fact, a large variety of different types of phase diagrams can be realized, because the interaction can be tuned *via* the size ratio between colloid and depletant, the polymer concentration and the types of repulsive interaction involved. This behaviour is further modified for polydisperse spheres [128]. As a consequence, this much more complex phase behaviour may significantly alter the crystallization pathways and involved kinetics and predictions of crystallization kinetics are much more difficult.

Experimental AS are mainly based on PMMA or micro-gel-PS HS systems with added polymer [11, 13, 37, 129, 130], silica spheres depleted with DNA in buffer solution [131], but also CS under low salt conditions depleted with polyethyleneoxide [132]. It has been demonstrated that the depletion efficiency of rods is even greater than that of coiled polymers [133]. Use of PnIPAM spheres as depletant seems to be particularly interesting because it allows a fast and precise *in situ* tuning of the interaction simply by temperature variation [134]. Also micelles, able to form and/dissolute under external control have been used as depletant of variable concentration, amongst others to study sublimation dynamics of AS crystals [135]. For some cases the pair potentials were directly measured in optical tweezing experiments or with total internal reflection microscopy [136, 137]. In general, a remarkable qualitative agreement is obtained between observed and predicted phase behaviour, given the even wider range of experimental control parameters as compared to HS and CS. Moreover, already early this century a series of experiments on gravitationally settling HS polymer mixtures demonstrated that there is an intimate connection between theoretically derived energy landscapes and the experimentally observed selection of phase transition pathways [138, 139, 140, 141]. The direct comparison of theoretical predictions for nucleation and/or growth rates and/or mechanisms to experimental data, however, is much harder due to the large parameter range to be observed in both approaches.

## 3. Instrumentation

### 3.1 Scattering

Due to its specific length and time scales, colloidal crystallization is conveniently studied from its very start up to the longest annealing times by optical methods like light scattering and microscopy. Melt structure and in particular short and long time dynamics are accessible from dynamic laser light scattering. Based on the assumption that a meta-stable melt is isotropic and homogeneous, mainly in-plane goniometer setups are used in combination with commercial correlator equipment. To deal with the problem of multiple scattering several cross-correlation experiments have been designed [142, 143, 144], and also heterodyning schemes have been proposed [145]. Exploiting high brilliance beam lines, also x-ray correlation spectroscopy is used more and more frequently [146, 147]. Crystal investigation is feasible by (time resolved) static light scattering experiments [4, 9]. This may be done by goniometer-mounted photo-detectors or in newer set-ups by line or area diode [148] or CCD detectors [149]. Statistics here are further enhanced by slowly turning the sample cell. To avoid possible sample disturbance, alternatively the sample may be fixed and the detector arm equipped with several detector arrays rotated as to average over the complete 2D-scattering pattern [150, 151, 152, 153]. The machine designed by Schätzel also featured another special feature. It allows for simultaneously collecting data in the Bragg and the small angle scattering regime, which yields complementary information also about the immediate environment of nucleating and growing crystals. 2D scattering patterns have also been collected employing flat or spherical screens [154, 155] or by direct imaging to a CCD camera [156]. A 2D back-scattering experiment using in addition multi-monochromatic illumination has recently been demonstrated [157].

Bragg light scattering is also exploited in measurements using reflection spectrometers. Here the scattering vector q varies with wavelength at fixed reflection angle. Devices are available commercially and have e.g. been used to study the gravitational compression dynamics in settling CS [158] While giving only one dimensional information on the Scherrer powder pattern, such machines are small and thus ideally suited for experiments under micro-gravity [159, 160]. A related technique adapted to colloids is based on Bragg diffraction interference fringes which were used to study salt dependent CS crystal growth [161]. Also transmission experiments on crystallizing colloidal suspensions [162] are meanwhile feasible with commercial UV-VIS spectrometers and useful in particular in experiments on wall nucleated crystals with preferred orientation [76, 132].

Even for systems matched in their refractive index to that of the solvent, increased turbidity and multiple scattering may become issues at elevated concentrations. In those cases, Small angle X-Ray or Neutron Scattering (SAXS or SANS) can be employed for polymer particles [163], and Ultra Small Angel X-ray Scattering (USAXS) can be employed for silica and other inorganic particles [164, 165, 166]. Extension to dynamical diffraction theory as done for small angle X-ray scattering should also be feasible there [167].Coherent X-ray imaging has been employed to characterize defect structures in fcc colloidal crystals [168] and recently also X-Ray microscopy has been used to investigate crystallizing samples of sedimenting, anisotropic inorganic colloids [169].

## 3.2 Microscopy

Advances in digital still and video imaging equipment as well as new developments in image processing and analysis based on enhanced computer power [170, 171, 172] have enormously boosted microscopy over the last two decades. Video microscopy combined with conventional and confocal microscopy came into extensive use to characterize colloidal interactions, structure and dynamics in colloidal fluids, glasses, gels and crystals as well as phase transition mechanisms and kinetics [173, 174, 175, 176, 177]. The resolution attainable with conventional microscopy even on low contrast samples is already impressive [178]. Confocal microscopy, however, allows to image and track colloidal particles in 3D [179] even in rather turbid samples and with high resolution and precise localization in x, y, and z direction [180, 181]. Resolution is further enhanced, if fluorescent particles or particles with fluorescent core are employed [182]. A very fascinating system resulted from dyeing the particles of binary mixtures in different colours [35] which allowed to study binary mixtures with unprecedented resolution. As in simulation, discrimination of disordered melt or glass regions from differently ordered crystalline regions is crucial [183, 184, 185, 186] and typically based on bond order parameter analysis, coordination number and Voronoi density determination both in 3D [34, 187, 188, 189, 190] and 2D systems [191].

Microscopy may be combined with a lot of other techniques. Quite fascinating is the possibility to combine it with static light scattering. Imaging of the back focal plane of the objective enables direct q-space analysis [192, 193]. Restricting the q-space used for image construction, the selective imaging of incoherently scattering point defects, dislocations and grain boundaries becomes feasible [194]. In addition, colour coding due to loss of Bragg scattered light allows fast large scale scans of the structure in (thin film) samples [195]. Illumination under Bragg condition allows to selectively monitor nucleation and growth of individual crystals or selected twin orientations [196, 115, 197]. Microscopy may further be combined with optical tweezers, e.g. to prescribe a guiding field for heterogeneous nucleation and/ or phase transitions at surfaces [198] and in the bulk [199].

## 2.3 Data acquisition and evaluation

Crystallite numbers and sizes, as well as their time derivatives, the nucleation rates and growth velocities are directly accessible in microscopy. In scattering experiments these have to be inferred from e.g. peak widths and integrated intensities measured as a function of time [149]. Most evaluations are based on KMJA theory [200, 201, 202, 203] assuming constant steady state nucleation rate densities and growth velocities. In experiments, one has access only to Bragg signals and images of crystals *after* these have formed, thus there is a minimum observable crystallite radius. Meanwhile, the density of the remaining melt might have changed, across coexistence by development towards the global equilibrium value, but also locally due to the formation of depletion zones about the initially compressed nucleus [204, 205]. This may alter the nucleation rate density in time, as has indeed been observed for CS at low meta-stability [206]. Steady state nucleation rate densities, $J$, as implicit in KMJA theory are observed only at low meta-stability. At larger meta-stability only time-averaged $J$ are accessible directly. Care has therefore to be taken to assure coincidence of data from different experimental techniques [207]. If $J$ is known as a function of time, one may also apply Kashiev's theory of transient nucleation to obtain an estimate of the steady state $J$ [208, 209].

To discriminate heterogeneous and homogeneous nucleation at the container wall in scattering experiments, evaluation has to exploit the 2D scattering pattern or individually address different peaks [210, 211]. To discriminate between these two scenarios in the bulk of a system, Turnbull devised a dispersion method isolating small volumes of melt to obtain a large number of independently monitored crystallization events [212]. However, variations in the size of the involved emulsion droplets may significantly alter the obtained nucleation rates [213]. A related technique employing micro-fluidics has recently been demonstrated by Gong et al. to study the nucleation of PnIPAM particles in isolated micro-fluidic droplets [214]. Microfluidic drop formation was also used to isolate small volumes of CS melts at low electrolyte concentrations, which were found to be quite stable against contamination with salt [215]. Use of this approach may, where necessary, eliminate the influence of nucleus-nucleus interactions, or for very small droplets allow studying the influence of confinement.

Finally, often also knowledge of the chemical potential difference, $\Delta\mu$, between the initial melt and the final crystal is necessary for further evaluation. For HS, $\Delta\mu$ is calculated for a given $\Phi$ from the known HS equations of state in the fluid and solid state [216, 217]. For other systems these equations of state can, in principle, be measured from sedimentation experiments [218] or measurements of the osmotic pressure [219]. In particular for CS, $\Delta\mu$ may also be derived from measured growth velocities and their dependence on interaction parameters [196, 115, 10].

## 4. Crystallization from a quiescent, homogeneous, meta-stable melt

### 4.1. Hard sphere homogeneous nucleation rates

Being Soft Matter, such suspensions are easily shear-molten by simply tumbling the sample or applying directed flow [220, 193, 221]. In the first case, the melt is typically assumed to be isotropic and homogeneous, in the second case, regions with broken symmetry and preferred orientation may remain. The first step in crystallization from an isotropic melt then is homogeneous nucleation followed by growth and coarsening. Within classical nucleation theory (CNT) [222, 223, 224, 225] the formation and growth of a spherical nucleus *via* single particle attachment is assumed to be an activated process in which the interfacial free energy of the nucleus-melt surface and the latent heat released upon phase transformation compete. The nucleation rate density, *J*, takes the form of an exponential with the exponent determined by the resulting barrier height and pre-factor by the particle number density, the limiting single particle attachment rate and the derivative of the Gibbs free energy curve at the barrier. This concept was originally proposed for atomic liquids condensing from a vapour phase but subsequently has been adapted to colloidal systems at various levels of sophistication. There are a number of issues of concern as to the applicability of this approach like the neglect of elastic stress and nucleus non-sphericity or the use of the capillarity approximation for very small nuclei [226, 227, 228]. Still, CNT is most widely used to interpret data obtained in nucleation experiments and simulations [14, 187, 207].

Alternatively, Dixit and Zukoski proposed a kinetic model which does not involve nucleation as an activated process [229, 230]. It is based on a depletion zone model similar to that of Derber [231] of a crystalline object forming and growing limited only by the interplay between the enhanced thermodynamic driving force and the reduced particle diffusivity in determining crystal nucleation rates as the particle density is increased. Scaled to the short time self diffusion coefficient, their model describes the nucleation rates, induction times and growth rates of experimental HS systems remarkably well without any further free parameters [230].

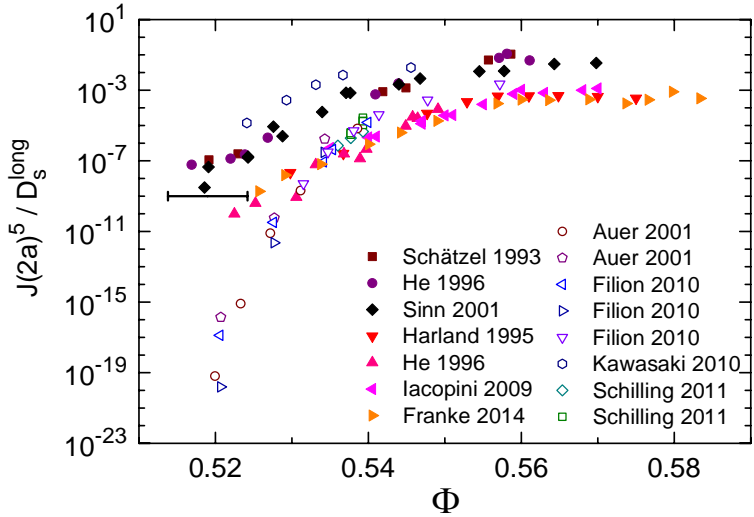

Fig. 1: scaled nucleation rate densities from experiment (closed symbols) and simulation (open symbols) in dependence on effective HS packing fraction. Data seemingly sort in tree groups. One are those obtained by Schätzel and Ackerson (PMMA, a = 500nm s = 0.05, in tetraline/decalin, squares) [204], He et al. (PMMA, a = 495nm, s = 0.05, in tetraline/decalin, dots) [232] and Sinn et al. (PMMA, a = 445nm, s = 0.025, in tetraline/decalin, diamonds) [233]. The second experimental data group comprises data of Harland et al. (PMMA, a = 495nm, s = 0.05, in tetraline/decalin, down triangles) [232, 149], He et al. (PMMA, a = 215nm, s = 0.07, in tetraline/decalin, up triangles) [232], Iacopini et al. (polystyrene micro-gels, cross-link density 1:10, a = 423nm, s = 0.065, in 1-EN, left triangles) [235] and Franke (polystyrene micro-gels, cross-link density 1:30, a = 410nm, s = 0.055, in 1-EN, right triangles) [236]. The horizontal error bar gives an estimate for an uncertainty in packing fraction of ±0.0051 in $\Phi$. The simulation data for HS systems form another group. Here we show data of Auer and Frenkel for monodisperse HS (circles) [237] and HS with s = 0.05 (pentagons) [238], of Filion et al. [239, 240] for monodisperse HS obtained from a Bennett–Chandler type theory where the nucleation barrier is determined using umbrella sampling simulations (left triangle), by forward flux sampling (right triangle) and by molecular dynamics (down triangle) as well as of Schilling et al. [240] obtained for monodisperse HS by molecular dynamics simulations (diamonds) and Monte Carlo simulations (squares). These simulations of HS show a much stronger packing fraction dependence of the scaled nucleation rate density at low $\Phi$ than experimental data. At large $\Phi$ the agreement between simulation data and experimental data of the second group is quite remarkable. For comparison we also show the simulation data of Kawasaki et al. (hexagons) [189] for a monodisperse system of spheres with WCA interactions used to approximate HS interactions [242].

By 1999 most nucleation data had been taken on HS-systems. Fig. 1 shows as a first highlight of research on colloidal crystallization the elder data and those obtained since the millennium. Here we compare results for scaled nucleation rate densities, $J^* = J(2a)^5/D_S^{long}$, where a denotes the particle radius, and $D_S^{long}$ is the long time self diffusion coefficient. We compile data on HS from experiment [204, 232, 233, 234 149, 235, 236] and simulation [237, 238, 239, 240, 241, 189]. Several points are to be noticed. First the available data base has increased enormously since 1999 [10]. Now data are available for experimental PMMA spheres, slightly charged PMMA spheres and PS micro-gel spheres for over about 18 orders of magni-tude in J* and packing fractions in the range of $0.515 \leq \Phi \leq 0.585$. Thus the upper part of the coexistence regime as well as the crystalline phase up to the glass transition has been probed. Moreover, a similar amount of data over a very similar range is now available from simulations employing different techniques and potentials.

Second, the experimental data appear to sort in two groups. One has to take caution here, because of the large systematic uncertainties still remaining in the packing fraction determination. We therefore include an exemplary optimistic error bar in Fig. 1. Taking the apparent grouping seriously, one may ask for possible correlations with other sample properties. PMMA particles are observed in both groups, micro-gels only in the lower. Taking the case of the PMMA spheres investigated by He [232] the authors note that both systems were synthesized, coated and conditioned and characterized exactly the same way. Thus here the systematic error in $\Phi$ can be neglected. The smaller particles could have a slightly softer potential because their coating is more extended as compared to the particle radius. For a softer potential, however, a notably increased J* is expected, as can be seen from the simulations of Kawasaki et al. [189] who used particles with a Weeks Chandler Andersen-potential [242]. Also Gasser et al. found for slightly charged HS ($\Phi_F = 0.38$) [187] that their rates from confocal microscopy were larger than for HS particles and showed only little $\Phi$ dependence. By contrast, the smaller experimental HS show lower J*. A second possibility to slow nucleation is polydispersity, as e.g. seen by comparing the two data sets of Auer and Frenkel [237, 238]. In the experiments, the smaller PMMA species in fact has the larger polydispersity. However, the size similar large PMMA-species of Sinn [233] are found within the same group as the large PMMA spheres of He despite a factor of two difference in polydispersity.

Another difference between the two PMMA species, already pointed out by the authors, could be the influence of gravity. In fact, as was recently noted by Russo et al. [243], the two groups of data sort by their significantly different Peclet numbers, $Pe$, which characterize the ratio of diffusion to sedimentation time scales. The upper group shows $Pe \approx 0.3$, the lower group $Pe \approx 10^{-2}$. Also, the sedimentation free simulations agree considerably better with the lower group, while data extracted from simulations with sedimentation, performed by Russo et al. on particles with WCA interaction at $Pe \approx 0.3$, agree reasonably well with the upper group data at $\Phi \geq 0.555$. At low $\Phi$ the simulation data are significantly above the upper experimental group, which might, however also be an effect of the slightly soft potential used. The authors could further improve the agreement between their simulation and the upper experimental group data by using a two state model [243]. More interestingly, these data are some 10 orders of magnitude above the other sedimentation free simulation data. Thus, if one takes the difference between the two experimental data groups as significant, sedimentation seems to be a viable candidate for its origin, although the actual mechanism is not yet understood. This conjecture could be checked in further experiments of differently sized PMMA-spheres and computer simulations including an external gravitational field and possibly also addressing density fluctuations during sedimentation [244], e.g. driven by the different density of nuclei.

A third observation that can be made in Fig. 1 is that all curves show a characteristic shape. At large meta-stability above $\Phi_M$ the data saturate, indicating that nucleation is mainly transport controlled by the long time relaxations still possible at such large densities. Here we see a remarkable agreement of experiment and simulation. Across the coexistence region the data strongly depend on meta-stability. Here the dependence is considerably stronger in the simulations. In fact, the simulations from different techniques agree within say one or two orders of magnitude but fall below both experimental data groups by some six orders of magnitude at $\Phi \approx 0.52$. It appears remarkable that the onset of this persistently remaining discrepancy coincides with the necessity of eventually forming a system with two distinct densities. Sev-eral suggestions have been made as to explain this difference. One is the role of sedimentation, as already discussed above. A second possibility seems to be an influence of the solvent viscosity. Schilling et al. [245] reported recently an accelerating influence of an increased solvent viscosity. Its influence was barely visible for $\Phi \approx \Phi_M$ but became quite pronounced as $\Phi$ was lowered to $\Phi = 0.537$ [246]. This observation points to the right direction, but clearly more data at lower $\Phi$ are desired. It should, however, also be noted that the theoretical calculations by Dixit and Zukoski using a purely kinetic model [229, 230] were demonstrated to coincide with the experimental data of Sinn, if scaled to the short time self diffusion coefficient. So the nucleation rate densities in the lower coexistence region remain an interesting puzzle.

*4.2. Homogeneous nucleation rates for highly charged spheres*

As of 1999 no kinetic data on CS nucleation were available. Meanwhile about half a dozen thoroughly deionized, highly charged single component systems and several mixtures [118, 165, 208 209, 210, 247, 248, 249, 250] have been studied, mostly by time resolved static light scattering. One particular species (PnBAPS68) has been studied by different techniques to follow the nucleation behaviour over some seven orders of magnitude in J [251, 206, 207]. Fig. 2 compiles some data in terms of the measured nucleation rate densities, *J*, plotted versus the number density. Note, the shown *n* correspond to packing fractions of 0.0015 to 0.1, except for PTFE180 which under index and buoyancy match could be investigated up to 0.15. The buoyancy matched PTFE180 system shows an increased *J* as compared to the non-matched. The mixture was nucleating at somewhat larger rates. Data for pure species and for mixtures show a qualitatively very similar dependence on particle concentration. The initial strong increase of *J* with $\Phi$ resembles that of HS. The corresponding melts all show a still high diffusivity and no kinetic glass transition is encountered [252, 253]. The slowing of the increase in *J* at elevated concentration is therefore solely attributed to an increased interfacial free energy, $\gamma$[207].

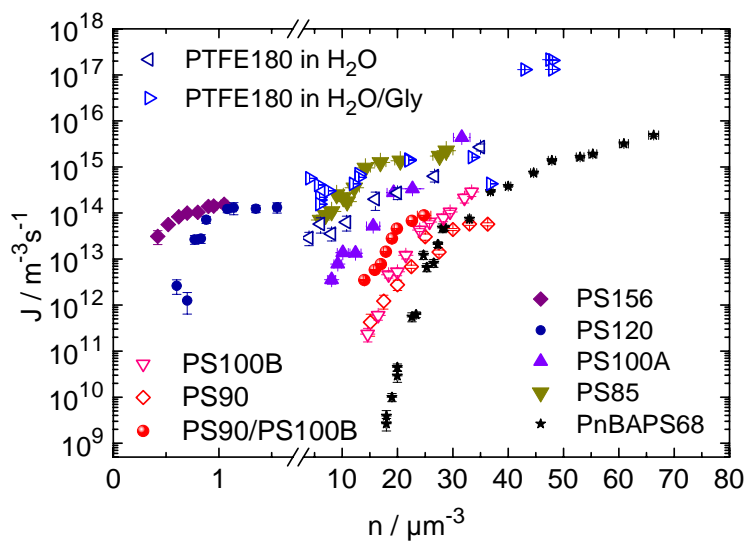

Fig. 2: Absolute nucleation rate densities obtained as a function of number density for several species of thoroughly deionized highly charged colloidal spheres and one mixture redrawn from [118, 207, 249 ]. In the quoted lab codes, PTFE denotes polytetrafluoroethylene particles, PS denotes polystyrene particles and PnBAPS denotes polystyrene-poly-n-butylacrylamide copolymer particles. The numbers give the particle diameters. Data on

PnBAPS68 (stars) were obtained from different techniques including microscopy all other data are from time resolved static light scattering [207]. PS100B (open down triangle) and PS90 (open diamond) were investigated both as pure species and in an equimolar mixture (balls) [249]. PTFE180 was investigated in pure water (left triangle) or in a buoyancy matching $H_2O$/glycerol mixture (right triangle) [118].

The crystallization of CS mapped to effective HS was studied theoretically with their kinetic model by Dixit and Zukoski [254]. Rates based on CNT were predicted by simulations [255]. Interpretation of experimental data, so far, mainly followed CNT. The reduced interfacial free energy, $\sigma = \gamma\, k_B T / d_{NN}^2 = \gamma\, k_B T\, n^{-2/3}$, i.e. this quantity normalized to the area taken by a single particle within the interface, is one of the key parameters of crystallization and of great interest also for metal physics and material processing [256]. It has been observed to be proportional to the molar enthalpy of fusion [257], and the proportionality constant, the so-called Turnbull coefficient has been determined from experiment and simulation for many fcc-crystallizing metals to be about 0.5. For bcc crystallizing metals only simulations are available, which suggest, that this quantity should be some 40% smaller [258, 259]. CS colloids show many similarities with metals and therefore provide an interesting test case. For both metals and CS, σ increases with increased meta-stability indicating an energetic contribution in addition to entropic contributions known from HS [260]. Very recently we have been able to extract both equilibrium reduced interfacial free energies, heats of fusion and Turnbull coefficients for several pure CS species and one mixture [261]. We observed a Turnbull coefficient of 0.31 in good agreement with the prediction from simulation.

For HS the nucleus-melt interfacial free energy was inferred from nucleation measurements in the frame of CNT [233] and the equilibrium crystal-fluid interfacial free energy was directly obtained from groove shape analysis of crystals at coexistence [262] and microscopic level investigations of the equilibrated interfacial structure and dynamics [187, 263] as well as from different theoretical approaches [264, 265, 266, 267, 268]. Observed values range between 0.51 $k_B T/(2a)^2$ and 0.66 $k_B T/(2a)^2$, agree reasonably between experiments and theory and show no systematic dependence on meta-stability. Values for slightly charged PnIPAM sphere crystal fluid interfaces are somewhat larger [269] but still well below those obtained for deionized CS. HS exhibit the theoretically expected Turnbull coefficient for an fcc crystallizing material, while HCY particles show both interfacial free energies and Turnbull coefficients much lower than metals and CS colloids [270]. Interestingly, Auer and Frenkel reported an increase in interfacial energy from crystallization studies on strongly polydisperse HS systems [238], however, without accounting for the possibility of fractionation [271]. By contrast, our recent study on CS indicates a decrease of σ with increased polydispersity [261]. Also in Fig. 2, the nucleation rate densities of the mixture (PS90/PS100B) are larger than those of the pure species, which, within CNT, corresponds to a decreased nucleus-melt interfacial free energy.

*4.3.     Homogeneous nucleation in other colloidal systems*

Systems with interaction types other than HS or CS have been measured less frequently and no systematic compilation can as yet be given. Crystallization in systems with depletion interaction is hard to catch, especially at elevated sphere packing fraction. Typically, a HS glass will melt upon addition of depletant but then vitrify again as an attractive glass [272, 273]. The range of observable crystallization is therefore restricted for both one component systems [44, 274] and binary mixtures [275]. Still and already very early, Dhont et al determined the crystallization rates for AS colloids [148, 276]. More recently, crystallization from isotropic attractive HS suspension with variable second virial coefficient, $B_2$, was studied by simulation and a change of the crystallization mechanism with increased packing fraction observed

[277]. The nucleation rates of PnIPAM particles have been measured by reflection spectroscopy by Tang et al. [76]. Interestingly, the authors observe that the crystal lattice constant shrinks at constant number density as the particle size is shrinking with increased T. This may indicate some attractive interactions being present. Unfortunately no values for $D_s^{long}$ are available to compare to J* in Fig. 1. If scaled by the Stoke-Einstein diffusion coefficient, the J* values at low $\Phi$ roughly coincide with the data observed for HS, show only little increase with packing fraction and therefore fall well below the HS data of Fig.1 at large $\Phi$. Finally, we note that the kinetic theory of Dixit and Zukoski has also been applied to the case of attractive model proteins [278].

## 5. Mechanism(s) of homogeneous nucleation

### 5.1 Monodisperse systems with various interaction types

As was shown, CNT is a useful tool for a relatively coarse-grained parameterization of the kinetics of nucleation and provides a handle to extract several key parameters of nucleation. On the other side, within its framework no data on the actual nucleation process can be obtained. Since typically, the crystal differs from its host phase in both density and structure, two extreme pathways are in principle possible: a densification of amorphous regions which subsequently attain the target structure, as well as a structural fluctuation towards the target structure which subsequently densifies. In addition, intermediate paths and intermediate (meta-stable) structures may appear [279, 280], a prominent example from metal physics being the crystallization of iron. Also, the final solid phase may display several polymorphs [281]. Furthermore, from density functional theory of Lennard-Jones particles, simulations on protein crystallization, and phase field crystal calculations the formation of densified precursor regions through critical fluctuations was suggested. Thus connections can be made to spinodal decomposition [282, 283, 284, 285, 286] and for very densely packed precursors also to frustration [287]. All these scenarios are far off the simplifications made in CNT. From a naive point of view, one might therefore expect HS to first densify then order, since immediately a large amount of entropy for the remaining melt is gained by compressing the precursors, which then later optimize their packing. For highly CS one might expect the opposite path to apply, since their crystallization is driven dominantly by enthalpic contributions which enforce an optimization of the local distances and configuration. While this provides a guideline for conceptualizing, hard work is currently going on in terms of quantitatively characterizing the paths actually taken.

To do so, methods with high spatial resolution are desired, but also light scattering can make some contribution. Having access to scattering from crystals only after their formation, light scattering nevertheless may also analyze changes of the scattering pattern at earlier times where it is specifically sensitive to density fluctuations. First indications of a two step nucleation mechanism in HS were indeed observed in time resolved static light scattering. [205]. Subsequently, also the influence of polydispersity and the kinetic glass transition were investigated on a series of similar PMMA particles [45, 288, 289] and in PS micro-gel particles [290]. In these studies densified amorphous precursors were formed, which exhibited no long range order and showed densification by several percent in packing fraction as compared to the initial melt density. Scattering patterns and the observation of ripening dominated growth suggested that the growing crystallites were embedded in the dense compressed region and hat this core shell structure also pertained during the growth stage [235, 291]. More detailed structural and dynamic information on the actual mechanism was gained in the following from confocal microscopy, e.g. on slightly charged HS [187] which showed nucleation into a close packed rhcp structure. The rhcp structure can be regarded as a random stacking of fcc and hcp planes and seems to be the kinetically preferred initial structure of HS nuclei [292], while on longer time scales, the structure transforms to fcc by annealing of stacking faults [293]

Also AS nucleate into this rhcp structure, when undergoing separation into regions of significantly different density. At very high compression, as possible e.g. for sticky HS and/or when the crystallization is coupled to spinodal decomposition [129, 218], crystallization may get stuck, if the attraction is too strong and/or the compression rate is too large. The terminal volume fraction of the structurally arrested state as a function of compression was e.g. studied by Li et al. [294] to find that sharply quenched HS systems are mixed of amorphous regions and a network of crystals. This can be rationalized by considering the random close packing limit proposed by Bernal and the structures formed at that packing fraction of $\Phi = 0.64$ [295]. Recent calculations by Anikeenko and Medvedev [296, 297] on highly compressed HS show the occurrence of a network of tetrahedral order, which is not suited to become space filling. Investigations in the coexistence range of an AS system showed less compressed crystals but an interesting mechanism for growth and ripening, which produced a power-law increase in the crystallite size, $L = t^\alpha$, with a common exponent close to one third. This value is expected for a conserved order parameter, which in this case could be identified as being the polymer concentration which had to equilibrate between crystalline and fluid ordered regions [298, 299]

Densified precursor regions were found in simulations by Schilling et al. on monodisperse HS [34, 241] and identified to nucleate into a random hexagonal close packed (rhcp) structure and to produce the observed scattering patterns of [205]. Lechner and Dellago observed nucleation occurring *via* solid clusters that comprise of a hcp core embedded within a cloud of surface particles that are highly correlated with their nearest neighbours but not ordered in a high-symmetry crystal structure [300]. Russo and Tanaka proposed from their simulations that bond order rather than density favours crystallization at low density HS but induces fivefold symmetry at large densities and thus suppresses crystallization [301]. Such a bond ordering would be expected to optimize the packing for more accessible free volume at equal packing fraction, while icosahedral packing would locally increase $\Phi$, and hence give more free volume to the surroundings. Five fold symmetric twinning of nuclei on the other side may result from the combination of stacking faults and point defects in densified regions [302].

Simulations on systems with softer interactions seem to favour bcc structures [303, 255] in accordance with the proposal of Alexander and McTague [280]. Moroni et al. studied a Lennard Jones system to find the probability for a nucleus to become a post-critical to be correlated not only to its size but also to its structure [304]. Russo and Tanaka studied a Gaussian core model in which they observe precursor regions of high bond-orientational order in which crystals of alike symmetry nucleate [305]. Formation of bcc crystals in the spinodal region was studied using a continuous Landau's free energy expansion and relaxation dynamics to find precursors of poly-tetrahedral structure embedded in local regions of icosahedral order. Bcc crystals then nucleate on the surface of the latter regions [306]. By contrast, Kawasaki suggests from observations on WCA particles, that nucleation is intimately connected to the medium range transient structural order in the liquid. Nucleation was found to preferentially occur in regions with high structural order via wetting effects [189]. From simulations on CS, Gu et al. found that long ranged interactions lead to bcc nuclei. At shorter ranged interaction, particles may get transiently trapped in amorphous clusters, which eventually nucleate to bcc crystals, or, if bypassed nucleate directly into the stable fcc phase [307].

Experiments on the nucleation mechanism of CS are still rare. A very recent confocal microscopy study by Tan et al. [190] compared nucleation in CS systems under variation of repulsion strength and range to nucleation in systems of slightly charged but screened PMMA HS at $\Phi \approx 0.53$. Thus, crystallization into both bcc and rhcp crystals was studied and analyzed in terms of bond order parameters, density from Voronoi analysis and coordination numbers. The authors report a remarkably complex scenario and some important differences to previous observations. For instance the precursor formation of the CS systems started in regions of less than average density, which increased in density to slightly above average upon ordering. Initially, for all systems several structures coexisted in the precursor regions with hcp dominating. Further structural development towards the (meta-stable) nucleus structures appeared to occur through a continuous collective transformation and therefore was controlled by the differing rates of different pathways.

Two-step nucleation during heterogeneous nucleation at cell walls – proceeding via a bcc structure, while the stable phase was fcc – was studied by Zhou et al. [308, 309]. On the other hand, shear may induce an fcc or rhcp phase, which is meta-stable with respect to both melt and final bcc crystal phase. The wall based fcc or rhcp layers then either decay completely or serve as extended nucleus for the stable bcc phase [310]. Two step nucleation was also observed in phase field crystal models [311] and experiments with driving fields being present, e. g. AC electric fields acting on sedimented CS [312]. In the latter case, the interesting observation made by Zhang and Liu was that the nucleation occurs through coalescence of pre-critical crystalline regions. Be it a continuous structural evolution or coalescence process, both, as cooperative processes, challenge the view of single particle attachment as traditionally proposed within CNT which possibly is more appropriate for liquid condensation from the vapour phase. The newly developed view might also contribute to an understanding of the long time stable amorphous states observed upon increasing the nucleation rate in CS systems. If precursors are formed by structural transformation of individual regions and, due to their abundance, intersect at different orientations, this would block further evolution by coalescence and keep the system quenched in its precursor state [118]. This kind of frustration would be on a higher organizational level than that usually discussed in the context of jamming and glass transition [287].

Comparing experimental data to CNT assumptions, we finally note, that also the idea of an initially isotropic melt, homogeneous in structure and dynamics, has been challenged by recent experiments. In confocal microscopy studies, Gasser et al. found the melt structure of HS to be mainly fcc, with some traces of icosahedral order which increase in amount with increasing $\Phi$ (3% at $\Phi = 0.06$) [188]. Taffs et al. observed 10mer clusters of fivefold symmetry to be the prevailing structure in slightly soft HS approximants [36]. Dominance of such structures was proposed as mechanism for dynamical arrest at large $\Phi$ [313]. Static structure factors measured by USAXS in CS suspensions at elevated concentrations in the meta-stable melt state show a split of the second peak, very similar to that observed in liquid Ni [314]. This feature was related by the authors to an icosahedral short range order [209]. Finally, high precision dynamic light scattering studies on PMMA [315] and on PS micro-gels [236] reveal, that only for the stable fluid the dynamics are Brownian, while additional modes appear in the meta-stable melt.

*Polydisperse systems and binary mixtures*

One may expect that the changed phase diagrams of polydisperse spheres [51, 52, 53, 54, 55, 128, 271] will influence the crystallization kinetics and mechanisms as compared to monodisperse case for larger values of the polydispersity, *s*. Already very early it was pointed out that there exists an empirical upper bound for crystallization to occur at about $s = 0.07$ [8]. Similar changes are therefore to be expected also for binary mixtures of small size ratio. In both cases compositional fluctuations are needed to form a critical nucleus in addition to density fluctuations. Monte Carlo simulations show, that the underlying phase diagram has a significant ef-fect on the mechanism of crystal nucleation: fractionation of the species upon crystallization increasingly suppresses crystallization, but under azeotropic conditions the nucleation barrier is comparable to pure fluids [316]. More specific, HS with a size ratio of 0.85 are a frequently employed model system in studies of meta-stable melts and the glass transition. On the one hand, their resistance to crystallization stems from their eutectic phase behaviour. The coexistence pressure is raised to $p_{co} = 22.5 k_B T/\sigma_P^3$. On the other hand, the nucleation barrier becomes extremely large in the vicinity of the eutectic composition, such that crystallization is also kinetically hindered [317]. When under such conditions the packing fraction is increased, the system vitrifies. However, Kozina et al. have demonstrated that it can be forced to crystallize upon adding some polymer and turning HS into AS [275]. The authors further observed very slow crystallization showing an inter-species fractionation dependent on initial composition but furthermore also signs of intra-species fractionation. Both PMMA and PS-micro-gel HS and their binary mixtures were used to investigate the effects of fractionation on nucleation pathway [318, 319] and kinetics [288, 289, 290]. In combination with Molecular Dynamics simulations, Pusey et al. presented an extensive analysis of previous crystallization experiments with the focus on both fractionation and glass transition [320].

Also growth should be modified by polydispersity, but again in different ways for HS and AS. For purely repulsive HS systems a diffusive depletion zone dynamically favours smaller particles to enter the crystal [321]. For HS with depletion attraction such a kinetic fractionation should be enhanced and modified by a gas layer forming between crystal and surrounding melt [322]. Sandomirski et al. have recently demonstrated that depletion zone formation and kinetics can also be studied experimentally in planar HS crystal melt interfaces [323], but systematic investigations of the influence of polydispersity are still under way. For the case of CS suspensions close to melting, it was observed that the reaction limited growth velocity decreased strongly when small amounts of a larger sphere species were added [324]. Upon adding more than some 15% (by number) of the larger sphere species the velocity decreased further but now only proportional to the averaged diffusion coefficient of the mixture. The growth velocity eventually returned to large values when only a small fraction of small particles was left. This was interpreted as a reduction of the initially large interfacial thickness to a monolayer interface upon contamination. It may be expected, that also for strongly polydisperse CS systems a similar mechanism will reduce the growth velocity.

Charged spheres show binary phase diagrams of spindle, azeotropic and eutectic type with dominant formation of substitutional alloy crystals [325]. Their crystallization dynamics are rather similar to those of the CS pure components and have been studied by light scattering and reflection spectroscopy [119]. Their sometimes very slow solidification may further be followed *via* the evolution of the elastic properties [326]. HS and AS, by contrast, show a rich phase behaviour including many compound structures known from atomic systems [327, 27, 35, 328, 329]. Alternating deposition of two differently sized CS species was demonstrated by Velikov et al. [330] and a similar technique applied to oppositely charged species by Tan et al. [331]. Both lead to layer by layer growth and the formation of compound structures with the formation kinetics controlled by evaporation and the suspension's surface tension. Crystallization of compounds in a bulk suspension containing mixture of oppositely charged spheres has been studied qualitatively by Sanz et al. in experiment and simulation [332]. However, systematic studies on the crystallization kinetics and the nucleation mechanisms involved are still missing.

## 6. Crystallization under external influences: manipulating the micro-structure

### 6.1 Heterogeneous nucleation

Heterogeneous nucleation is a well established means of generating desired micro-structures in metal systems. Examples are inoculation leading to grain refinement or templated growth to produce high quality single crystals [333]. A similar approach has been followed also in colloidal systems [374, 334]. Even more than for metals, colloidal crystal microstructure can be manipulated and controlled by templating, seeded growth and various external fields like gravity, electric fields, temperature and chemical gradients, shear fields and combinations thereof.

By contrast to metals it is here also possible to study nucleation at walls which are flat on the scale of a single particle, but of course also substrates structured on length scales equal to [335] or larger than the typical particle distances [131]. Since hard spheres wet a hard wall [336, 337], heterogeneous nucleation can be studied on flat horizontal [338] and flat vertical walls [151]. The heterogeneous nucleation rates of HS at a hard and CS at a charged vertical wall were measured in dependence on meta-stability of the melt to observe a wetting transition in the latter case [211, 248]. As a result, the competition between heterogeneous and homogeneous nucleation becomes biased and the resulting micro-structures differ qualitatively [210]. A straightforward experiment to grow large scale columnar crystals is the sedimentation of hard spheres, where the growth velocity, and hence the quality of the crystals, is controlled by the density mismatch and can be described by Kynch theory [339].

Templated growth is easily possible and has been demonstrated already very early in the seminal paper by Tang et al. [335]. The wetting behaviour of crystals on differently structured substrates has recently been reviewed in [340]. From simulations on HS, it was observed that there is a pre-freezing transition on a suitably patterned substrate, the lattice mismatch decides whether crystallization occurs *via* wetting or heterogeneous nucleation, and that a substrate with square symmetry possibly can stabilize the meta-stable bcc phase [341, 342, 343, 344, 345]. Experimentally, heterogeneous nucleation can be further assisted by applying additional external fields during growth. Perfect single crystals with NaCl structure were grown from binary HS mixtures settling in electric, gravitational and dielectrophoretic fields to a structured surface [346]. The boundary conditions for successful colloidal epitaxy under gravitational settling were investigated by Hoogenboom et al. [347, 348, 349, 350, 351] using both 2D crystalline and 1D line templates. Centrifugation onto templates was studied by Jensen et al. [352] and also the nucleation mechanism on 111 fcc templates [353], which differs from that observed for flat walls [338]. Still different and more complex are the observations made for AS on structured and strained substrates [354]

Seeding has been investigated with seeds of various sizes with either smooth or structured surfaces. At least two factors determine whether a large impurity with smooth surface can function as a seed for heterogeneous nucleation: timescales and surface curvature [355]. At large seeds an interesting kinetic mechanism was observed by Sandomirski et al. in which the crystal first nucleates at the seed surface but then detaches and from the bulk grows backwards [356]. From simulations, such a "catalytic" nucleation was expected only for seeds just above the critical size for assisting nucleation [357]. From the experiments on flat templates and also theoretically [358] it is expected, that a seed of structure and lattice constant commensurate to the stable crystal phase will enhance nucleation considerably. Templating with bulk seeds of fcc structure made of sintered silica spheres was studied by Schöpe et al. [359] under µ-gravity. They observed a lot of subtle dependencies in their study and conclude that in addition to seed concentration, size and lattice mismatch, also the degree of meta-stability exerts significant influence on the kinetics. Recently also seeding without material seeds was demonstrated with optical tweezers [360]. A corresponding simulation inserted a free floating finite sized template into the simulation box to find, that the template particle motion should not exceed the Lindemann limit for inducing crystallization [361]. Local nucleation without actual seed was also realized *via* laser induced local depletion interaction [362]. Seeding in 2D was demonstrated with local electro-osmotic pumping [363, 364].

Often, not all seeds are actually used for nucleation. Remaining seeds then are either distributed throughout the sample or populate the grain boundaries of the emerging crystals. Their accumulation there is assisted by annealing processes [81, 365] and a foam-like microstructure appears similar to that observed for long lasting annealing of AS crystals [298, 299]. Cleaning of contaminated crystals can be achieved by processes analogous to zone melting [366].

It is also possible to poise crystallization by impurities. Villeneuve et al studied its dependence on impurity size in HS systems with large impurities forming the centre of grain boundaries. Crystal growth was inhibited to a greater extent near smaller impurities, pointing to local crystal frustration induced by the curvature of the impurity [367]. Recently, the transition from poisoning to enhancing nucleation has been studied quantitatively in a CS model [368, 369]. The resulting crystal size first increased then rapidly decreased as the inoculants' concentration was increased.

*External fields*

Since soft matter by definition is mechanically weak, it may be easily manipulated by external fields [370, 371]. There has been a strongly growing interest, to exploit this also crystallization and micro-structure control [372, 373, 374]. A rather large body of studies have been reported and only few examples can be given here to illustrate the richness of mechanisms and effects. Gravity, e.g. is often used to sort particles by size [375] and induce crystallization only in certain regions formed by stratification [243, 376]. This is also a means to control the competition between glass formation and crystallization [377]. In combination with heterogeneous nucleation at the bottom cell wall both components of a eutectic could be brought to crystallize even below the eutectic pressure [313]. There a concerted growth of columnar crystals of each particle species was observed. Gravity also acts on already growing crystals: dendrites, occurring under μ-gravity conditions [378] are sheared off [379]. Columnar growth is observed also upon relaxation of a strongly compressed sediment [380], in dielectric bottles [381] (sometimes combined with additional field induced transitions [382]), in direct or indirect chemical gradients of various kinds [99, 383]. An important observation there, is the feedback mechanism for diffusio-phoretic transport within the applied gradient. Hence, not only the electrolyte concentration is altered, but also the particles are redistributed, which may be mistaken for an equilibrium phase transition [120, 121]. Interestingly, however, in slowly moving gradients the actual interface displays an equilibrium characteristic [384]. Combining gradients with gravity, very complex morphologies can be produced. Yamanaka et al. [385] dropped NaOH into a suspension of silica spheres and observed most fascinating microstructures due to the interplay of gravitational settling of first the drop, then of the growing crystallites (compacted by diffusio-phoretic transport within the NaOH gradient) and finally the diffusive upward distribution of the NaOH. The final microstructure consisted of giant column-like crystals growing on a polycrystalline stratum. Complex morphologies may also be observed when one mechanically drives a polycrystalline material up a stationary gradient of electrolyte concentration [386].

Finally we mention the possibility of using shear to manipulate the microstructure as already demonstrated in the eighties [220]. Several sophisticated experiments have been devised to study phase transitions under shear [387, 388, 193]. If shear is applied during nucleation, the homogeneous nucleation rate is decreased and this leads to an increase in size of the final crystals [389]. Also bimodal crystallite distributions can be obtained by applying the nucleation suppressing shear only for limited amounts of time [9]. Further, a preferred orientation of homogeneous nuclei was observed in simulations [390]. Even small shear rates suffice to suppress dendrite growth [379]. Shear may further induce and orient heterogeneous nuclei yielding anisotropic micro-structures [155, 308, 309, 388, 391, 392, 393]. Finally, shear processes may also provide compartments for crystallization in micro-fluidic devices [394]. Due to limited space, we could review only a handful of examples in this novel and rapidly expanding focus of colloidal crystallization. These may suffice to illustrate the numerous possibilities to steer crystallization. The particular interest arisen in this kind of studies, however, also was facilitated by the qualitative and quantitative understanding of the basics of colloidal crystallization, which was achieved in the years before

**Conclusion**

The last fifteen years have seen a rapid expansion of our knowledge and understanding of colloidal crystallization kinetics and mechanisms. In this enterprise, experiments, simulations and theoretical approaches appeared to have continuously complemented each other. While this seems to be trivial, the importance of beholding the differences between the approaches cannot be overestimated. Colloidal spheres are perfect objects only for theory and simulation, while in experiments much care has to be taken as to carefully characterize the investigated particles, interactions and boundary conditions. With this knowledge at hand, theory and simulation always have to exercise a certain awareness of the level of coarse graining taken and of the kinds of approximations made. Moreover, also the optical experiments employed provide important complementarities. Light scattering experiments are capable of delivering data from large sample volumes with high statistical accuracy and reliability, their interpretation, however sometimes faces difficulties due to the loss of phase information and thus spatial resolution. Microscopic methods are extremely well suited for studies of crystallization mechanisms and structural details and their distribution. Statistical uncertainty due to the restricted sample volumes studied can here be overcome by repeating the experiments after carefully restoring the initial conditions. Therefore, in comparing experimental data to expectations from theory and simulation and vice versa, a further crucial aspect is the statistical accuracy in each of the compared approaches.

Concluding, we have learned a great lot from individual clever work within each of those approaches. But even more we have often understood from the fruitful controversy in comparing and discussing results obtained in complementary studies. The review has also shown that the insights now gained on the more fundamental aspects already proved very valuable for the processing of colloidal materials.

**Acknowledgement**

It is a great pleasure to thank E. Bartsch, M. Heinen, D. M. Herlach, J. Horbach, H. Löwen, W. van Megen, T. Okubo, H. J. Schöpe, and P. Wette for the numerous fruitful discussions we had. Financial support was granted by the DFG (Projects Pa459/13, 16, 17 and He1601/24) and is gratefully acknowledged.